\documentclass[
aps,%
12pt,%
final,%
notitlepage,%
oneside,%
onecolumn,%
nobibnotes,%
nofootinbib,%
superscriptaddress,%
noshowpacs,%
centertags]%
{revtex4}
\usepackage[english]{babel}
\usepackage[pdftex]{graphicx}
\usepackage{graphicx}
\usepackage{subfig}
\usepackage{float}
\usepackage{color}
\usepackage{tabularx}
\usepackage{threeparttable}

\def\lesssim{\mathrel{\hbox{\rlap{\hbox{\lower4pt\hbox{$\sim$}}}\hbox{$<$}}}}
\def\rs{{\rlap.}^{\prime \prime}}
\def\rl{{\rlap.}^{s}}

\begin{document}

\title{\textbf{{H$_2$O  MASERS and PROTOPLANETARY DISK  DYNAMICS \\ in IC~1396~N
\let\thefootnote\relax\footnote{Accepted for publication in Astronomy Reports, 2017} 
}}}

\author{\firstname{O. S.} \surname{Bayandina\footnote[1]{Contact e-mail: bayandina@asc.rssi.ru}}}
\affiliation{\scriptsize
Astro Space Center, Lebedev Physical Institute, Russian
Academy of Sciences, Moscow, Russia}

\author{\firstname{I. E.} \surname{Val'tts}}
\affiliation{\scriptsize
Astro Space Center, Lebedev Physical Institute, Russian
Academy of Sciences, Moscow, Russia}

\author{\firstname{S. E.} \surname{Kurtz}}
\affiliation{\scriptsize
Instituto de Radioastronom\'\i a y Astrof\'\i sica,
Universidad Nacional Aut\'onoma de M\'exico, Morelia, M\'exico}

\author{\firstname{G. M.} \surname{Rudnitskij}}
\affiliation{\scriptsize
Moscow State
University, Sternberg Astronomical Institute, Moscow, Russia}

\author{\firstname{A. V.} \surname{Alakoz}}
\affiliation{\scriptsize
Astro Space Center, Lebedev Physical Institute, Russian
Academy of Sciences, Moscow, Russia}

\begin{abstract}
\vskip 0.5mm
\footnotesize{We report H$_2$O maser line observations of the bright-rimmed globule
IC~1396~N using a ground-space interferometer with the 10-m
RadioAstron radio telescope as the space-based element. The source was
not detected on projected baselines $>$2.3 Earth diameters, which
indicates a lower limit on the maser size of $L>$0.03~AU and an upper
limit on the brightness temperature of
6.25$\times$10$^{12}$~K. Positions and flux densities of maser spots
were determined by fringe rate mapping.  Multiple low-velocity
features from $-$4.5~km/s to $+$0.7~km/s are seen, and two
high-velocity features of V$_{LSR}$$=$$-$9.4~km/s and
V$_{LSR}$$=$$+$4.4~km/s are found at projected distances of 157~AU and
70~AU, respectively, from the strongest low-velocity feature at
V$_{LSR}$$=$$\sim$0.3~km/s.  Maser components from the central part of
the spectrum fall into four velocity groups but into three spatial
groups.
Three spatial groups of low-velocity features detected in the
2014 observations are arranged in a linear  structure
about $\sim$200~AU in length.  Two of these groups were not
detected in 1996 and possibly are jets which
formed between 1996 and 2014.  The putative jet seems to
have changed direction in 18 years, which we explain by the
precession of the jet under the influence of the gravity of
material surrounding the globule. The jet collimation can
be provided by a circumstellar protoplanetary disk. There is a
straight line orientation in the V$_{LSR}$-Right Ascension diagram
between the jet and the maser group at
V$_{LSR}$$=$$\sim$0.3~km/s.  However, the central group with the
same position but at the velocity V$_{LSR}$$\sim$$-$3.4~km/s falls
on a straight line between two high-velocity components detected in 2014.
Comparison of the low-velocity positions from 2014 and 1996,
based on the same diagram, shows that the majority of the masers maintain
their positions near the central velocity
V$_{LSR}$$=$$\sim$0.3~km/s during the 18 year period.}
\end{abstract}
\maketitle

\newpage
\section{INTRODUCTION}

\subsection{General information}

IC~1396~N is a cometary, bright-rimmed globule in the northern
part of the HII~region IC~1396 located at a distance of 750~pc
\cite{matthews}.  Within the globule is an intermediate-mass
protostar IRAS~21391$+$5802. It is the brightest source among the 25
IRAS point sources embedded in the nebula, with a far infrared
luminosity of 440~L$_{\odot}$ \cite{fuente}. Multiple young
stellar objects (YSO) at different evolutionary stages are located
in the globule, and have been detected by their millimeter
continuum emission \cite{beltran1, codella}.  The BIMA-2 molecular
core, the most prominent source at 3.1~mm, is the center of
several bipolar molecular flows observed in wings of many thermal
lines (see \cite{sugitani, wilking, fuente, beltran1} and
references therein). 

\subsection{Masers}
H$_2$O maser emission was first reported by
\cite{gyulbudaghian}. Although IC~1396~N/BIMA-2 has been 
searched for OH emission various times, no such masers have been found.
OH emission in IC~1396~N was not found with the Nan\c{c}ay radio telescope (France) in
2002-2003 \cite{edris}.   We also observed IC~1396~N in the OH lines at 1665 and 1667~MHz
with the Nan\c{c}ay radio telescope in 2007 with null results at a 3$\sigma$ level of
0.1~Jy.  Methanol class~II maser emission at 6.7~GHz \cite{slysh67}, commonly found in accretion disks,
is not present, but there is Class I methanol maser emission at 44~GHz \cite{ic44ghz}, which indicates
a very early stage of the compression of protostellar matter. 
IC~1396~N was not covered by any of the \textit{MSX} or
\textit{Spitzer} surveys; no information is available about possible
association with Infrared Dark Clouds \cite{irdc} or Spitzer Dark Clouds
\cite{sdc} (IRDC/SDC) or with Extended Green Objects (EGO) \cite{ego}.

\subsection{VLA and VLBA studies}
In 1992 IC~1396~N was mapped in the H$_2$O maser line using the VLA with an
angular resolution of 0.1$^{\prime\prime}$ \cite{tofani}: they found
a  group of three maser spots within $0\rs18$  (135 AU) of one another
and a separate  component at a distance of $13\rs3$  (10$^4$ AU).  
Their monitoring program showed that the H$_2$O masers in IC~1396~N are variable in
both velocity and flux density.

In 1996 the H$_2$O masers in IC~1396~N were observed with the VLBA as
part of the prelaunch VSOP survey; the angular resolution was of order
0.3~mas \cite{migenes}.  A more complete analysis of the VSOP survey
data was presented by \cite{slysh}.  Their image, made with a
1.0$\times$0.3~mas restoring beam, shows an aligned chain of maser
spots about 15~AU in extent.  The strongest maser component, at
V$_{LSR}$$=$0.5~km/s, was unresolved on the shortest baseline, with a
correlated flux density of about 150~Jy: this was the strongest H$_2$O maser flux
density ever observed from IC~1396~N since its discovery in 1988
\cite{gyulbudaghian}. The VLBA observations also revealed two
high-velocity features at $-$14.1~km/s and
at 9.3~km/s were observed at distances of 410~AU and 10,000~AU from
the center of the cluster, respectively.

Three models were proposed by \cite{slysh}:  $i$) a Keplerian disk,
$ii$) a stellar wind, and $iii$) a molecular outflow.
A combination of these models was also discussed by \cite{slysh}. They 
suggest that the
low-velocity features arise in the Keplerian disk with
maser emission excited by shock waves traveling in the disk, while the
high-velocity features arise at the root of the molecular outflow
originating from the central 4~M$_{\odot}$ young star or
protostar. The mass of the disk and its angular momentum are similar
to those of the Solar System planets. It is suggested that it is a
circumstellar accretion disk accumulating the excess angular momentum
of the collapsing molecular core, which may give rise to the formation
of a planetary system.

In \cite{patel} the H$_2$O masers were observed with both single-dish
telescopes and interferometers.  Their VLBA observations had a synthesized
beam of 0.8$\times$0.4~mas (0.6$\times$0.3~AU).  These observations were
made in 1996 March, April, and May, just before the observations of 
\cite{migenes,slysh}, but the paper \cite{patel} was published
later than \cite{slysh}. The authors found a loop of maser spots within 1~AU
and a few higher velocity components moving away from the loop; the average proper
motion of the masers was 2~mas~per~year. 
They interpreted this picture as a bipolar outflow rather
than a large-scale disk, though the disk model could be valid for
the features closest to the central star. It was concluded that
additional observations may still reveal some rotation or even
``infall''\  toward the source. In \cite{patel} a strong variability
of the maser in 1995$-$1996 was also observed.

Of particular interest is the model proposed in \cite{rudnitskij1,
rudnitskij2, berulis}, in which  ``curved''\ and ``twisted''\
filaments of H$_2$O masers can be found, formed as a result of the
circumstellar protoplanetary disk precession. This model assumes
that a small gas-dust disk is surrounded by a massive near-star
torus, which remains from the protostar clusters and from which a
new star is forming. The axis of the disk is inclined at some
angle to the torus, making the disc precess. The outflow is
collimated by the disc and carries away material from the internal
walls of the torus, exciting masers where the material strikes the
torus walls. A large-scale bipolar flow is directed through the
polar regions of the torus. An example of this model is the
spiral nebula NGC~2261 (the Hubble Nebula)
\cite{lightfoot}.

Using the model of the precessing disk, along with the estimated size and
mass of the system, it can be shown that the precession period is
of the order of decades. Under the influence of the shock wave,
maser spots' proper motion can reach 1$-$2~mas~per~year and may be
traced over short time intervals in observations with a resolution
of (0.1$-$0.01)~mas.

The relatively close distance of  this source as well as the presence of
several YSOs, outflows, and possibly a protoplanetary disk or
protostellar wind, make it an ideal laboratory for the study of the
general aspects of the process of star formation.

We have monitored the H$_2$O  masers in IC~1396~N with the
RT-22 radio telescope in Pushchino (Moscow region) for more than
twenty years. Although these observations cannot resolve the masers'
spatial structure, they do provide a long history of the spectral
behavior. Our observations show that the peak flux density can
reach values of 100~Jy, while the spectral features appear and
disappear at different radial velocities. These fluctuations mean
that any study of the maser spots' proper motions should be done on
relatively short timescales to ensure the unique identification
of each maser spot being followed.

Positions of maser spots in multiple epochs can be obtained to a high
accuracy from VLBI observations with baselines larger than the
diameter of the Earth, thus tracing proper motions on relatively short
timescales.  With this goal, an experiment has been made using the
spaceborne RadioAstron telescope and a network of ground-based radio
telescopes.  The results are reported in this paper. In these
observations proper motions can be detected on timescales of a few
months, allowing us to determine the spatial structure of the masers
and to trace the dynamics of the maser spots for comparison with the
models discussed in \cite{slysh, patel}.  Furthermore, observations
with high spatial resolution using RadioaAstron can provide
information about the velocity field on scales $<$1~AU around the
central YSO.

\section{OBSERVATIONS}

We observed the H$_2$O masers of IC~1396~N
in the period from July to December~2014 with the
ground-space interferometer including the 10-m RadioAstron space
telescope and a network of ground-based telescopes. RadioAstron was
launched into orbit on July~18, 2011, from the Baikonur cosmodrome
by a Zenit-3F rocket, using a Fregat-SB booster. This is
the first ground-space radio interferometer operating in four radio bands
from meter to centimeter wavelengths; it provides the
highest angular resolution ever achieved of about 10~$\mu$arcsec.  The main
interferometer parameters are described in \cite{kardashev} and
presented in detail on the project website at
\url{http://www.asc.rssi.ru/radioastron/index.html}.

The observations were performed at a frequency of 22.2280~GHz with
a frequency resolution of $\sim$7.81~kHz, i.e., a velocity resolution
of 0.11~km/s, at the coordinates of the source
RA(2000)$=$21$^h$40$^m$41$\rl$75,
DEC(2000)$=$$+$58$^\circ$16$^\prime$11$\rs$9. For the
RadioAstron Space Telescope at a wavelength of 1.35~cm the System
Equivalent Flux Density (SEFD) is 37~kJy \cite{kov14}.

\begin{table}[H]
\scriptsize
\begin{center}
\caption{Baseline length and effective resolution}
\begin{tabular}{clccccc}
\hline \hline

Session &\multicolumn{1}{c}{Month} & Average    &Average    &Average         & Effective       &~~~Effective  \\
number &      &~~~projected~~~ & ~~~projected~~~ & ~~~projected~~~ & resolution      &~~~resolution \\
       &      & baseline   &baseline   &baseline        &                 &          \\
       &      & ED     & km        &10$^9$$\lambda$ &~~~10$^{-9}$ radian~~~& mas \\
\hline

1 & July    & 2.3 & 29\,310 & 2.2 & 0.46 & 95 \\
2 & October & 3.8 & 48\,420 & 3.6 & 0.28 & 58 \\
3 & November& 5.9 & 75\,178 & 5.6 & 0.18 & 37 \\
4 & December& 5.1 & 64\,984 & 4.8 & 0.21 & 43 \\
5 & December& 3.8 & 48\,420 & 3.6 & 0.28 & 58 \\

\hline
\end{tabular}
\end{center}
\end{table}

IC~1396~N was observed in five sessions, that fall into three
categories according to the size of the ground-space baselines: 
  short baselines up to $\sim$2.3~Earth diameters (July), medium
  baselines of $\sim$3.8~Earth diameter (one session in October and
one in December), and long baselines of
$\sim$5.9~and~$\sim$5.1~Earth diameters (in November and December,
respectively). In Table~1 we list the
average projected baseline lengths and angular resolutions for each of
the five observing sessions.  In Table~2 we list
the date and time for each observing session, the participating radio telescopes
and the peak flux density observed for the V$_{LSR}$$=$$-$9.4 km/s maser component.

\begin{table}[H]
\scriptsize
\begin{center}
\caption{Radio telescopes participating in H$_2$O maser
observations of IC~1396~N in~2014}
\bigskip
\begin{tabular}{ccccccc}
\hline \hline

Session~~& Date & Session  & Radio$^{\dag}$      & Session   &~~Session~~& Flux$^{\ddag}$ \\
number~~~&      &~~~duration~~~& telescopes &~~beginning~~& end      & ~~Density~~ \\
        &      & hours    &            &           &          & Jy        \\
\hline

1 & 2014-07-26 & 3 & Ys, Nt, Sr, Tr, Kl, RA & 04:00:00 & 07:00:00 &  4.8$\pm$0.2 \\
2 & 2014-10-10 & 4 & Ys, Kl, Ef, Tr, RA     & 20:30:00 & 00:30:00 & 12.2$\pm$0.8 \\
3 & 2014-11-23 & 1 & Ef, Tr, Kl, Sv, RA     & 05:00:00 & 06:00:00 & 20.2$\pm$0.8 \\
4 & 2014-12-01 & 1 & Ef, Sr, Kl, RA         & 21:00:00 & 22:00:00 & 51.3$\pm$0.8 \\
5 & 2014-12-10 & 1 & Sr, Tr, Sv, Zc, RA     & 10:00:00 & 11:00:00 & 28.4$\pm$0.8 \\

\hline
\end{tabular}
\end{center}

$^{\dag}$ RA $-$ space radio telescope RadioAstron (international project); Ef
$-$ Effelsberg  (Germany), 100-m; Sr $-$ Sardinia (Italy), 65-m; Kl $-$
Kalyazin, Moscow Region (Russia), 64-m; Ys $-$ Yebes (Spain), 40-m;
Nt $-$ Noto, Sicily (Italy), 32-m; Tr $-$ Toru\'n (Poland), 32-m; Sv $-$
Svetloye, Leningrad Region (Russia), 32-m; Sc $-$ Zelenchuk,
Karachay-Cherkess Republic (Russia), 32-m.

$^{\ddag}$ Calibrated flux density of  the strongest spectral feature at the
velocity  V$_{LSR}$$=$$-$9.4~km/s is given only for individual
telescopes $-$ see explanation in the Discussion section.

\end{table}

\begin{figure}[H]
\centering
\includegraphics[width=0.8\linewidth]{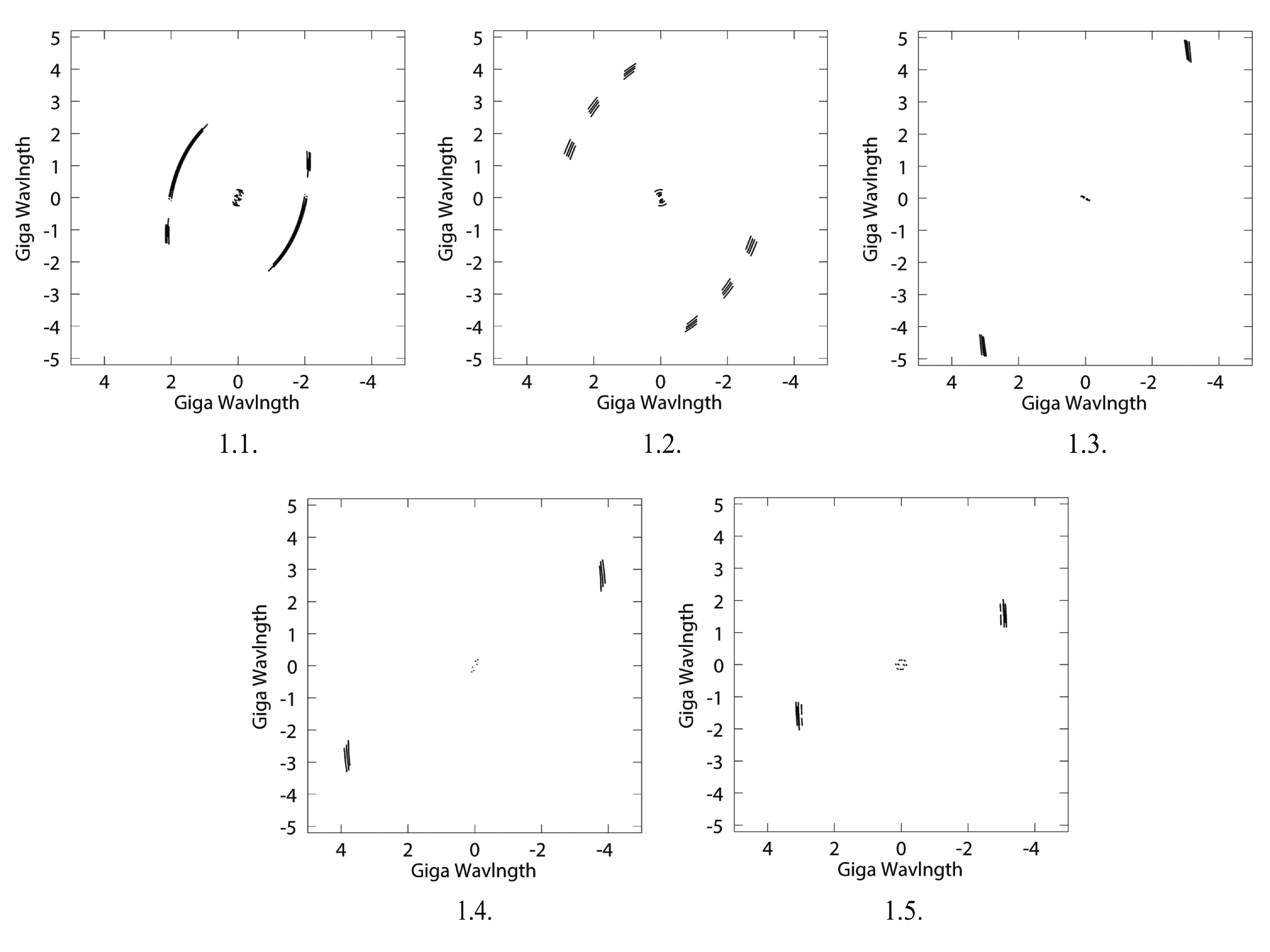}
\caption{\small{The $uv$-plane coverage for the
ground-space baselines (the data for all telescopes participating in the observations): 
1.1 $-$ July,
short baselines, session duration is 3~hours (session number 1 in Tables 1 and 2); 1.2 $-$ October,
medium baselines, 4~hours (session number 2); 1.3 $-$ December, medium baselines,
1~hour (session number 3); 1.4. $-$ November, long baselines, 1~hour (session number 4); 1.5. $-$
December, long baselines,  1~hour (session number 5).}}
\end{figure}

Observations on the short baselines provide the highest probability of signal detection, 
and allow us 
to establish a lower limit on the maser size
and an upper limit on the maser brightness temperature. The inclusion of medium
baselines improves the spatial resolution and image quality compared to that
obtained with only the short baselines.  The longest baselines were included
to permit the possible measurement of the maser spot size.

Plots of the ground-space $uv$ coverage for each observing session are
shown in Fig.~1.1$-$1.5 (the data shown are
for all telescopes participating in the observations). In
Fig.~2.1$-$2.5 we show the $uv$ coverage of
the ground-only baselines (the data shown are for only one baseline in each
session, on which the signal was detected $-$ the baselines shown are identified in Fig. 5).

\begin{figure}[H]
\centering
\includegraphics[width=0.8\linewidth]{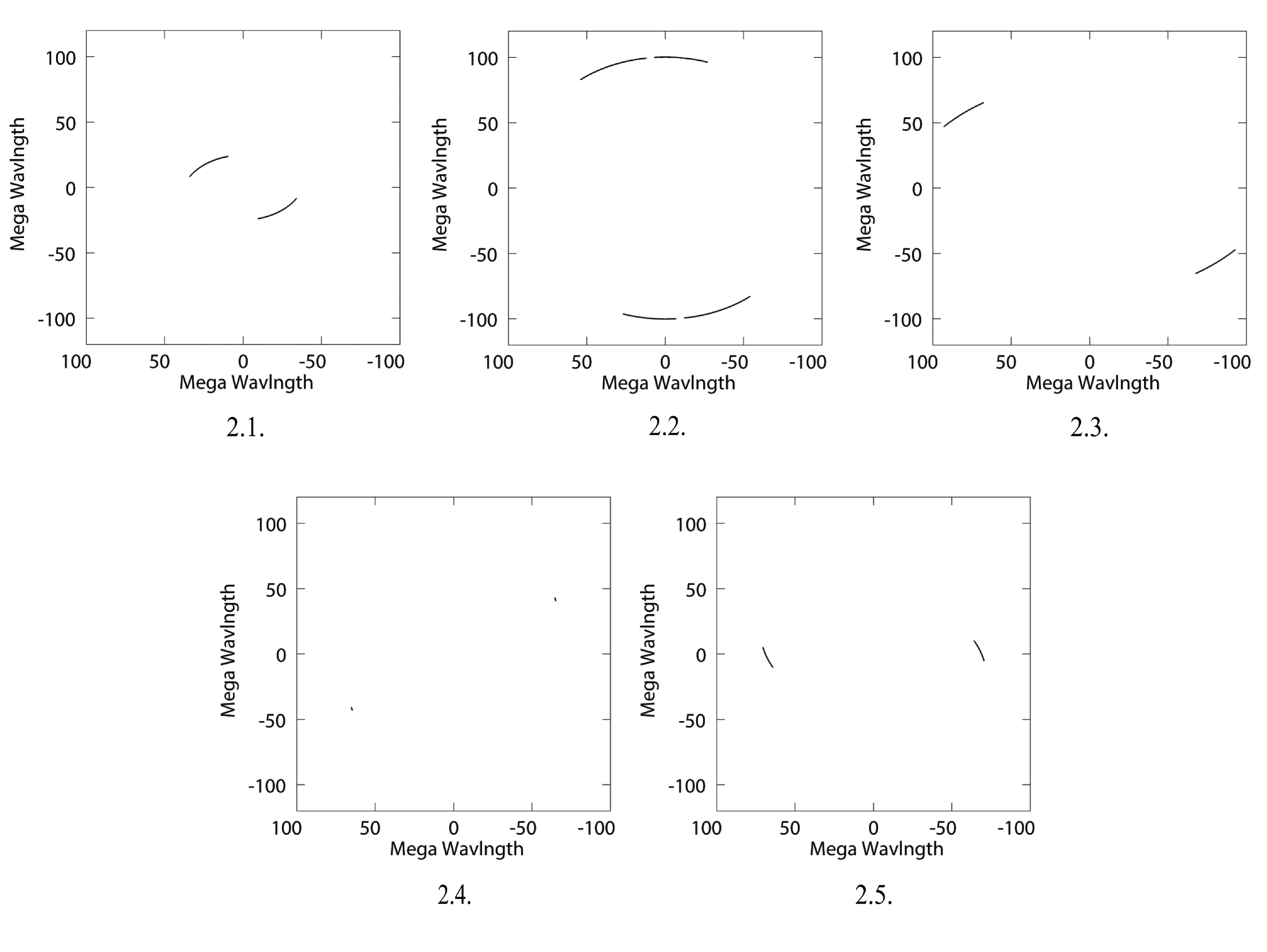}
\caption{\small{Same as in Fig.~1, but for ground-ground baselines (data are shown for only one baseline in each session, on which the signal was detected $-$ the corresponding baselines for each session are indicated  in Fig. 5).}}
\end{figure}

\section{DATA REDUCTION AND PRESENTATION of the RESULTS}

Primary data processing was carried out with the software correlator
at the Astro Space Center~LPI \cite{andr14}. The search for fringes
on the  ground-space baselines was performed using the software
package PIMA (\url {http://astrogeo.org/pima/}). Subsequent processing of
FITS files was done with
the AIPS software package (\url {http://www.aips.nrao.edu}).
Amplitude calibration was performed with the task ANTAB,
incorporating the antenna noise
temperature measurements made at the time of the observations.
Due to technical problems, recording of calibration tables at
telescopes Ys, Sr and Nt was not performed; in these cases we used
standard values of the noise temperature calculated on the basis of
known antenna parameters
(\url{http://www.evlbi.org/user_guide/EVNstatus.txt}).

\begin{figure}[H]
\centering
\includegraphics[width=0.77\linewidth]{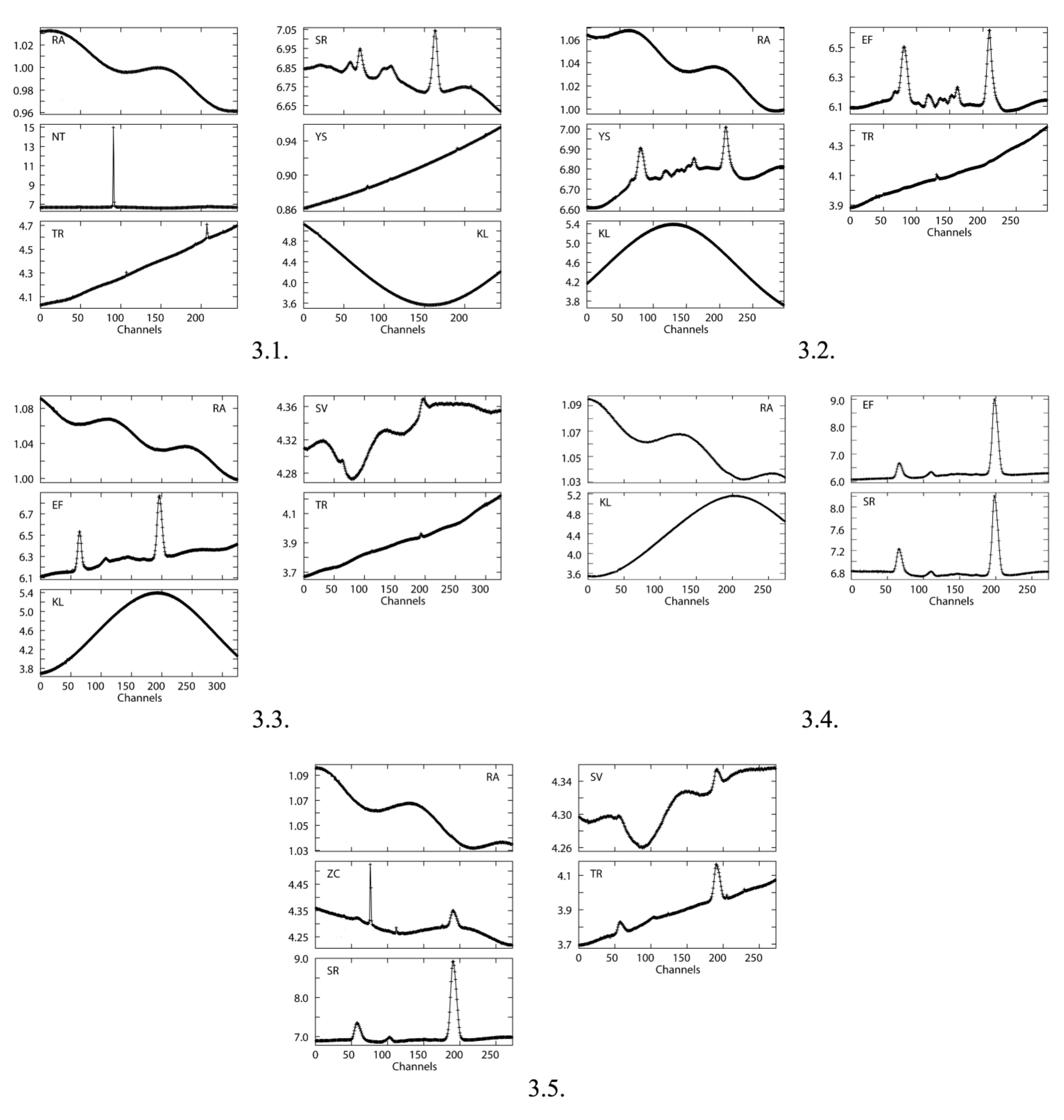}
\caption{\small{Initial uncalibrated auto-correlation spectra for all
telescopes participating in the 5~observing sessions
(3.1$-$3.5). For a key to the telescope abbreviations, see Notes to
Table~2.}} 
\end{figure}

The quasar J2137$+$510
(RA $=$ 21$^h$37$^m$01$\rl$0015, DEC $=$ $+$51$^\circ$01$^\prime$36$\rs$079)
was used as the calibration source, and was observed for five minutes in
each observing session at
all ground telescopes. The primary calibration of the group delay
signal and phase rate were performed with the task FRING, using J2137$+$510.
Bandpass calibration was done using the same quasar.  The absolute J2000 coordinates
RA $=$ 21$^h$40$^m$41$\rl$81, DEC $=$ $+$58$^\circ$16$^\prime$12$\rs$0 of the brightest
spectral feature at V$_{LSR}$$=$$-$9.4~km/s were determined from fringe-rate mapping.
Self-calibration was performed using the same maser feature.

\begin{figure}[H]
\centering
\includegraphics[width=1\linewidth]{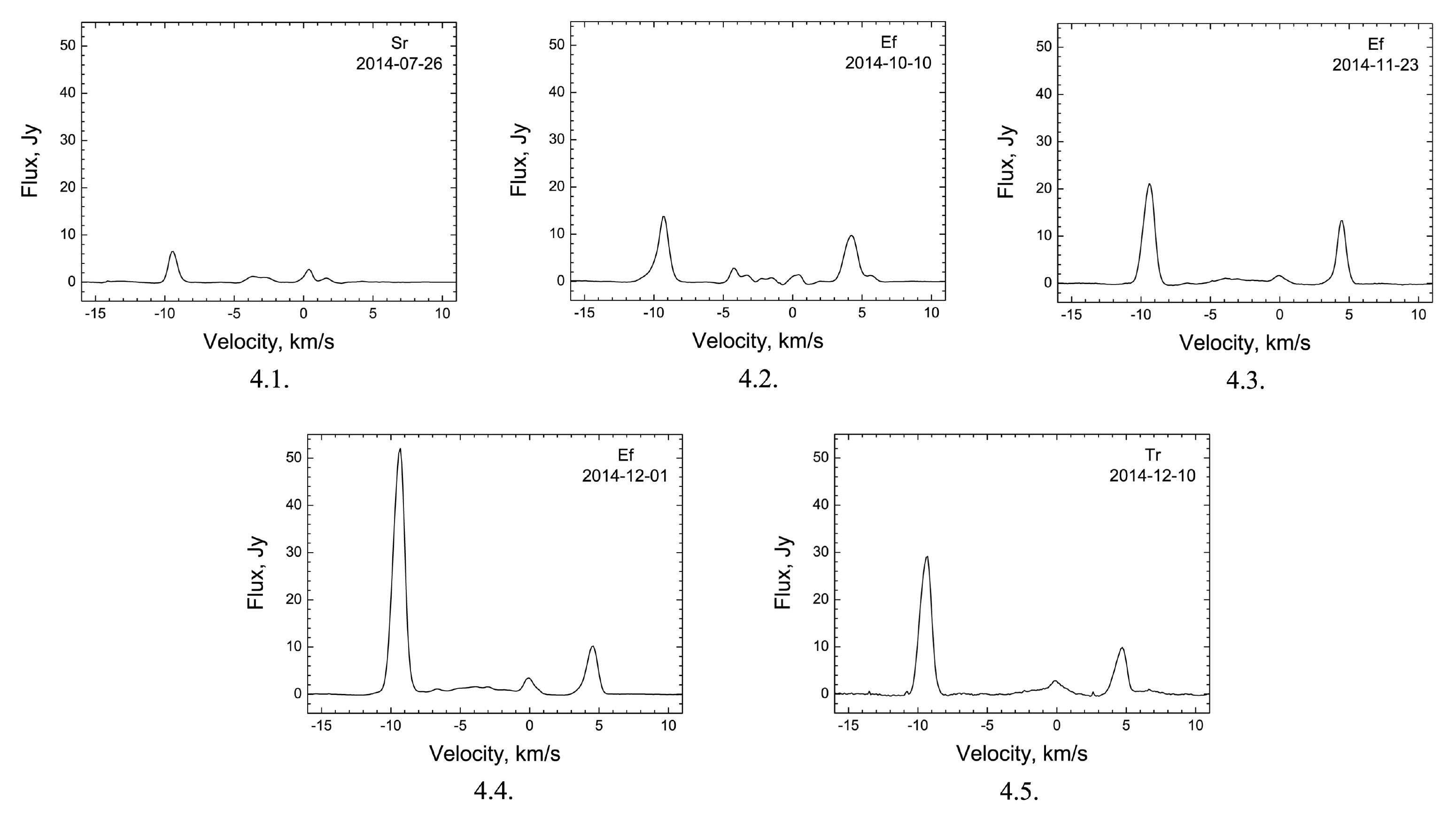}
\caption{\small{Calibrated auto-correlation spectra for individual
telescopes that were used to determine the source flux density in
each session of the observations. The telescope and the date are
indicated in the upper right corner.}}
\end{figure}

Fig.~3.1$-$3.5 present
uncalibrated auto-correlation spectra for all participating
telescopes, grouped by observation date.
Fig.~4.1$-$4.5 present the calibrated
auto-correlation spectra for the individual
telescopes that were used to determine the flux density of the
source for each session.
Fig.~5.1$-$5.5 present calibrated
cross-correlation spectra for only one baseline in each session on which the signal was detected. All spectra were constructed using the
task POSSM, and the calibrated spectra were scaled using SETJY.

\begin{figure}[H]
\centering
\includegraphics[width=1\linewidth]{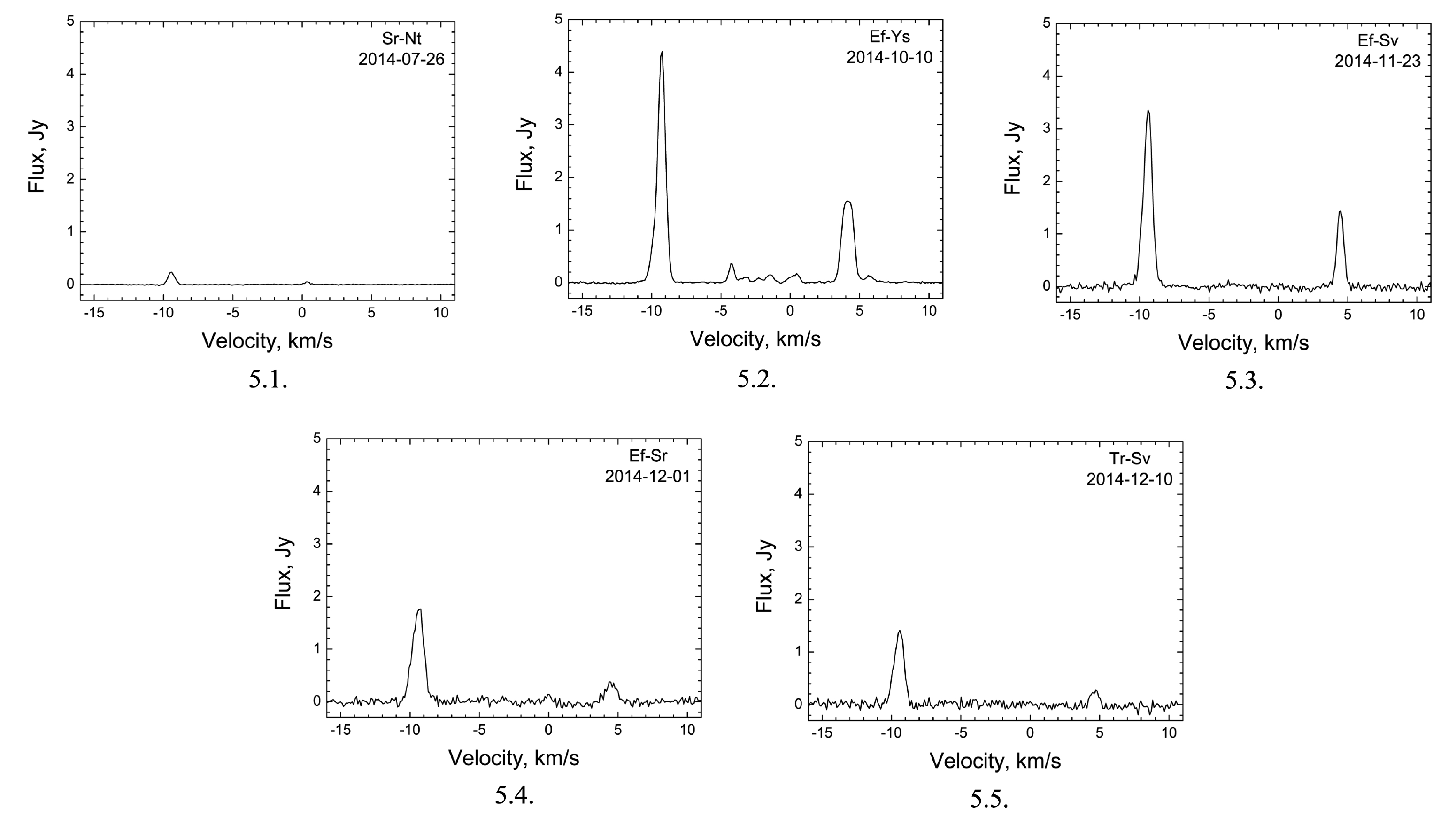}
\caption{\small{Calibrated cross-correlation spectra (Stokes $I$) for only one baseline (indicated in the upper right corner) in each session for which the signal was detected. The $Y$ axis
is the flux density in Jy; the $X$ axis is the line-of-sight velocity
(V$_{LSR}$ in km/s).}}
\end{figure}

Positions and flux densities of maser spots relative to the reference
spectral feature (V$_{LSR}$$=$$-$9.4~km/s) were determined by fringe
rate mapping (task FRMAP). All observing sessions had a reliable
detection (on the ground baselines) of the two high-velocity maser
features at V$_{LSR}$$=$$-$9.4~km/s and V$_{LSR}$$=$4.4~km/s.  For the
low-velocity features, only the second epoch (October) resulted in
detections at the 4$\sigma$ level or higher.

In Fig.~6 we present fringe rate maps for the
low-velocity features observed in the October 2014 session, covering
multiple components from about $-$4.6 to $+$1 km/s.  These
features roughly separate into four groups consisting of maser spots
nearby to one another on the sky; see Table~3 for their coordinates
and the Discussion section for a description of the four
groups.  In Table~3 we list the derived positional offset on a
channel-by-channel basis for those channels that had emission above
a 4$\sigma$ level.   The spectrum corresponding to the Table~3 data
is shown in Fig.~7.1$-$7.2.

\begin{figure}[H]
\centering
\includegraphics[width=0.7\linewidth]{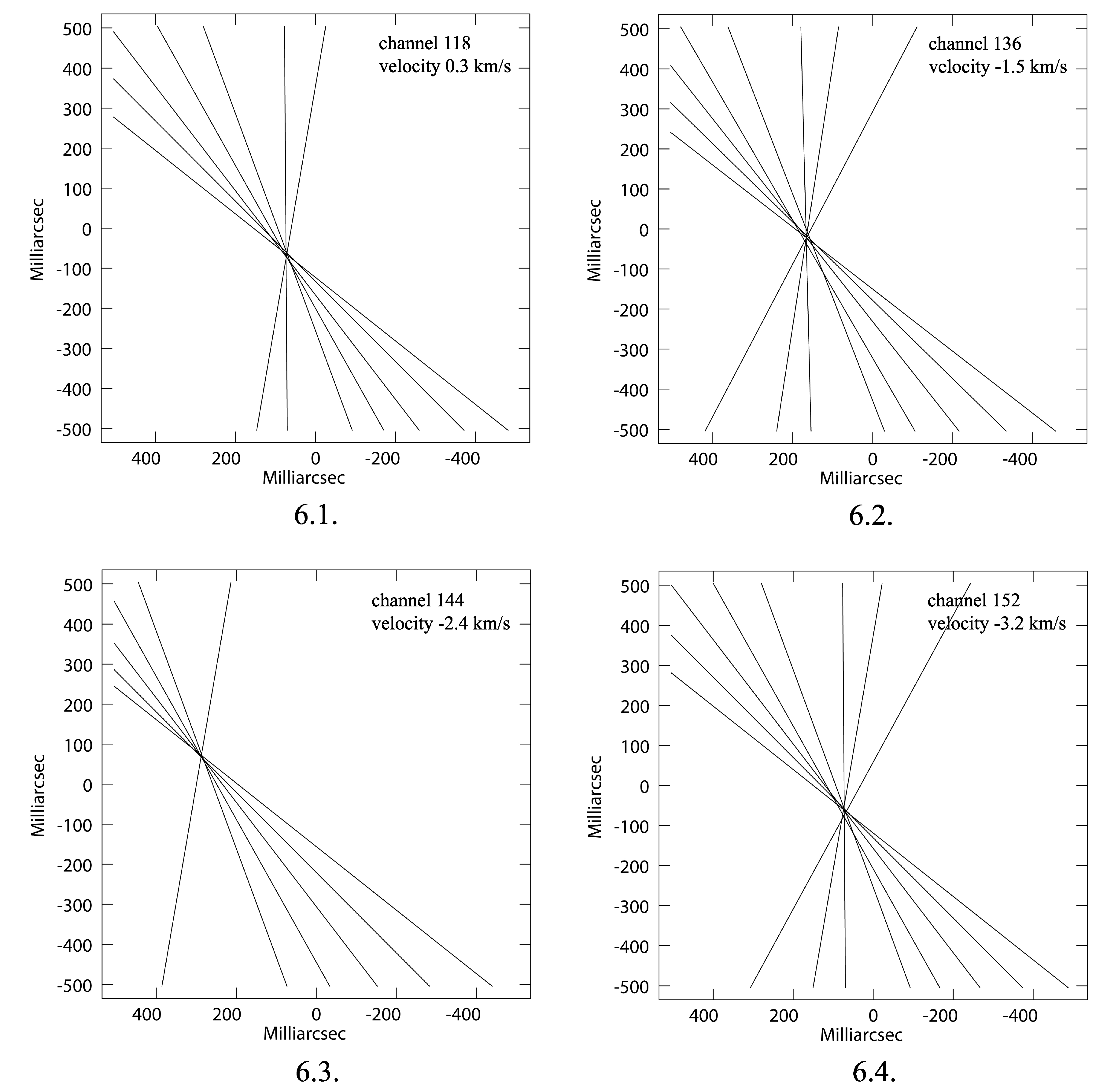}
\caption{\small{Maps determined by fringe rate mapping (task FRMAP) for
the brightest spectral  features in each of the four low-velocity groups observed in the second epoch (October
2014); see Table~3 and  the
description of the velocity groups in the Discussion section. The
channel number and the corresponding velocity are given in the
upper right corner of each panel.}}
\end{figure}

\begin{figure}[H]
\centering
\includegraphics[width=0.7\linewidth]{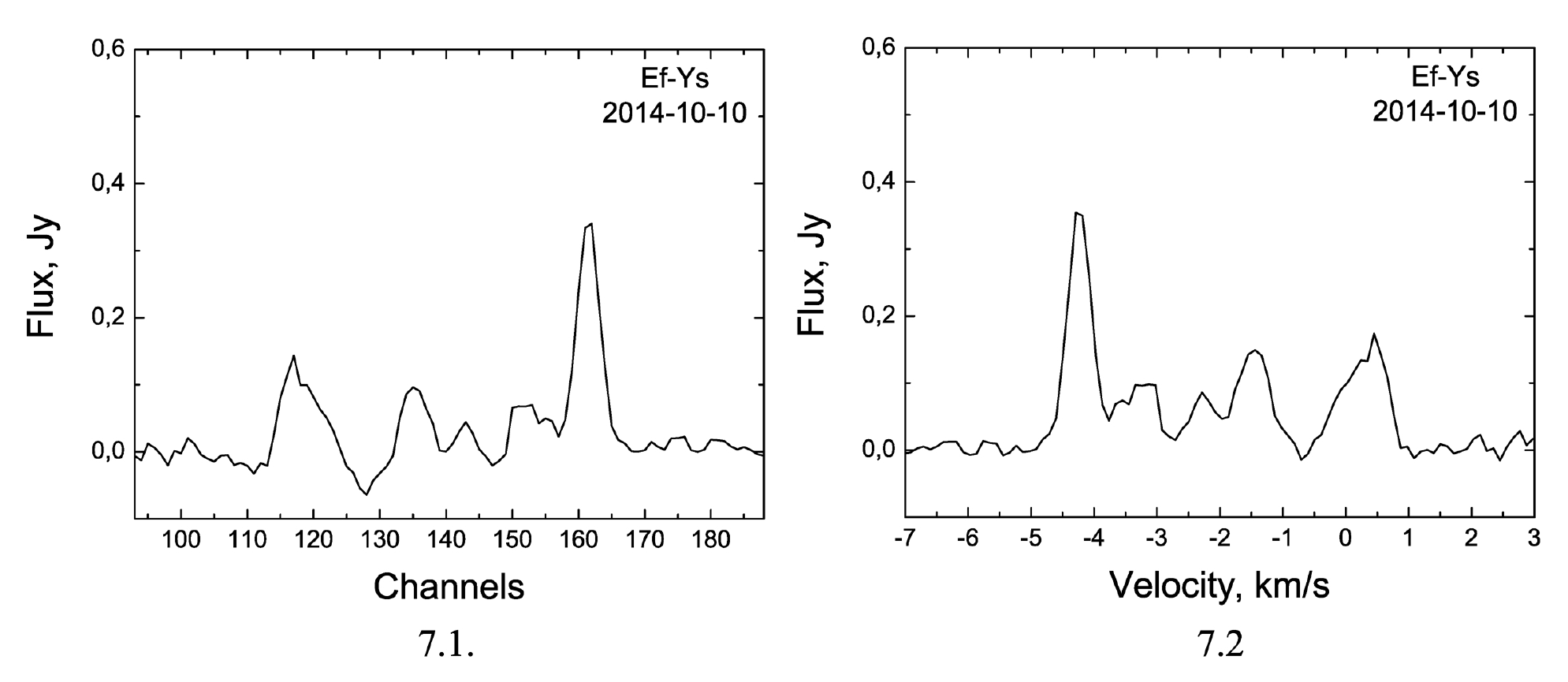} 
\caption{\small{The central part of the cross-correlation spectra for the October
session of the 2014~observations. On the $Y$ axis is the calibrated
flux density; on the $X$ axis at the left (7.1) are channels, at the right (7.2) is
line-of-sight velocity.}}
\end{figure}

\clearpage
\begin{table}[H]
\scriptsize
\begin{center}
\caption{List of the derived positional offsets on a channel-by-channel basis for channels from the central part of H$_2$O spectrum of IC~1396~N obtained in the October 2014 session (duration
4~hours)$^{\dag}$.}

\bigskip
\begin{tabular}{ccccccc}
\hline \hline

Spectral & Spectral  & Channel & V$_{LSR}$& $\Delta$RA & $\Delta$Dec& Flux    \\
feature  & component & number  &          &            &            & density \\
number   & number    &         &          &            &            &         \\
         &           &         & km/s     &  mas       &     mas    &  Jy    \\
\hline

1&1    &115&    0.7& 75.9$\pm$1.5&  $-$59.1$\pm$2.1 &   0.3\\
2&     &116&    0.6& 85.6$\pm$2.2&  $-$50.8$\pm$2.9 &   0.5\\
3&     &117&    0.5& 74.9$\pm$1.4&  $-$59.2$\pm$2.0 &   0.7\\
4&     &118&    0.3& 79.0$\pm$0.6&  $-$49.6$\pm$0.9 &   0.8\\
5&     &119&    0.2& 76.7$\pm$0.3&  $-$58.8$\pm$0.4 &   0.7\\
6&     &120&    0.1& 87.4$\pm$1.2&  $-$55.9$\pm$1.8 &   0.6\\
7&     &121&    0.0& 84.1$\pm$1.0&  $-$61.6$\pm$1.6 &   0.5\\
8&     &122& $-$0.1& 81.6$\pm$0.5&  $-$49.4$\pm$0.9 &   0.5\\
9&     &123& $-$0.2& 82.9$\pm$0.8&  $-$50.7$\pm$1.2 &   0.5\\
10&    &124& $-$0.3& 81.0$\pm$0.4&  $-$57.9$\pm$0.6 &   0.4\\
&&&&&&\\
11&2   &133& $-$1.2& 163.1$\pm$1.3& $-$23.5$\pm$1.8 &   0.5\\
12&    &134& $-$1.3& 163.0$\pm$0.2& $-$14.6$\pm$0.3 &   0.5\\
13&    &135& $-$1.4& 166.1$\pm$0.3& $-$14.8$\pm$0.5 &   0.6\\
14&    &136& $-$1.5& 170.6$\pm$1.4& $-$13.8$\pm$1.8 &   0.7\\
15&    &137& $-$1.7& 170.0$\pm$1.0& $-$17.5$\pm$1.4 &   0.6\\
16&    &138& $-$1.8& 173.2$\pm$0.4& $-$11.3$\pm$0.5 &   0.5\\
17&    &139& $-$1.9& 175.6$\pm$0.5& $-$13.2$\pm$0.7 &   0.4\\
&&&&&&\\
18&3   &142& $-$2.1& 297.5$\pm$1.2&    79.3$\pm$1.6 &   0.3\\
19&    &143& $-$2.3& 293.2$\pm$0.8&    85.4$\pm$1.2 &   0.4\\
20&    &144& $-$2.4& 292.8$\pm$0.3&    77.8$\pm$0.3 &   0.5\\
21&    &145& $-$2.5& 288.4$\pm$1.3&    76.5$\pm$1.9 &   0.4\\
&&&&&&\\
22&4   &150& $-$3.0&  72.8$\pm$0.5& $-$52.6$\pm$0.6 &   0.4\\
23&    &151& $-$3.1&  73.0$\pm$1.3& $-$55.6$\pm$1.4 &   0.5\\
24&    &152& $-$3.2&  74.7$\pm$1.1& $-$54.6$\pm$1.4 &   0.6\\
25&    &153& $-$3.3&  79.4$\pm$0.7& $-$43.5$\pm$0.8 &   0.5\\
26&    &154& $-$3.4&  71.8$\pm$0.4& $-$59.1$\pm$0.5 &   0.5\\
27&    &155& $-$3.6&  75.9$\pm$0.5& $-$53.5$\pm$0.7 &   0.4\\
28&    &156& $-$3.7&  83.2$\pm$1.3& $-$48.1$\pm$1.6 &   0.4\\
29&    &163& $-$4.4&  84.1$\pm$0.8& $-$43.8$\pm$0.9 &   0.3\\
30&    &164& $-$4.5&  75.9$\pm$5.3& $-$56.1$\pm$7.6 &   0.3\\

\hline
\end{tabular}
\end{center}
\begin{flushleft}
$^{\dag}$ The reference high-velocity
feature ($\Delta$RA$=$0, $\Delta$Dec$=$0) is at
V$_{LSR}$$=$$-$9.4~km/s, flux density 5.5~Jy. The position of other the high-velocity feature
at V$_{LSR}$$=$$+$4.4~km/s with flux density 1.7~Jy is $\Delta$RA$=$153.2, $\Delta$Dec$=$146.3.
\end{flushleft}
\end{table}

\newpage
\section{RESULTS OF THE ANALYSIS}

Fringes from  IC~1396~N on the ground-space
baselines were not detected in any of the observing
sessions. This could have two causes: either the Radio Astron
sensitivity was insufficient or the source is larger
than can be detected with such high resolution.  We explore
these two options below.

First, we consider the flux calibration of the data, and the
resulting measured values for the maser flux densities.
The Effelsberg 100-m telescope participated in three observing sessions 
(the second, third, and fourth).  Moreover, for this
telescope we have ANTAB tables; thus an accurate flux
calibration is possible for these three sessions.  
In the first session, the Sardinia 65-m radio telescope was
the only telescope with a detection.  It lacked an ANTAB table, so
the default SEFD was used.  In the fifth session, several telescopes
detected emission, with Sardinia being the largest of these.  Nevertheless, we
based the calibration on the spectrum of the Toru\'n 32-m telescope, because 
it had an ANTAB table, so its flux density could be determined more reliably 
than by using the default SEFD of the Sardinia telescope.

In Table~2 the flux density of the strongest spectral feature at the
velocity of $-$9.4~km/s (derived from the auto-correlation spectra, see Fig. 4) is presented for each session. These values
were determined using the ORIGIN package
(\url{http://www.originlab.com/}) by a procedure of the Fit Gaussian program.

From the auto-correlation spectra presented in Fig.~4, strong source
variability is evident.  The strongest flux density  of $\sim$51~Jy was during
the fourth observing session  (Fig.~4.4); the weakest flux density
($\sim$5~Jy) was in the first session (Fig.~4.1).

If we adopt the 5 Jy flux density as a lower limit, then the
lack of correlations with the space-telescope indicates that
the source brightness temperature at the time of the
observations was lower than 6.25~$\times$10$^{12}$~K.  The lack
of correlation on the shortest baseline of 2.3~ED means that the
linear source size is $L>0.03$~AU.

In each session of the observations maser emission was detected on a single
ground-ground baseline, as indicated in Fig. 5.  The telescopes not involved in the
detection were flagged before further processing with FRMAP.
In these detections, H$_2$O maser emission at the 4$\sigma$ level
 is present over a range of velocities  from about
$-$10~km/s to $+$5~km/s. There is no high-velocity blue feature
at V$_{LSR}$$=$$-$14.1~km/s \cite{slysh}, or at
V$_{LSR}$$=$$-$14.6~km/s \cite{patel}.  Nor is there a high-velocity
red feature at V$_{LSR}$$=$9.3~km/s, as observed by
\cite{slysh}. Rather, there are two new and bright features: a blue
one at V$_{LSR}$$=$$-$9.4~km/s and a red one at V$_{LSR}$$=$4.4~km/s.

The central, low-velocity  part of the spectrum  is much fainter
than in the observations reported by \cite{slysh}, but it covers a broader
range from V$_{LSR}$$=$$-$4.5~km/s  to V$_{LSR}$$=$0.7~km/s. The
feature at  1.32~km/s, used for the self-calibration
in \cite{slysh}, is not present in our spectra.

\section{DISCUSSION}
\subsection{Analysis of the maps and position-velocity diagrams}

Fig.~8  presents a map of the H$_2$O maser
spots for the central part of the spectrum in the velocity range
$-$4.5~km/s to $+$0.7~km/s and also of the two high-velocity
features at V$_{LSR}$$=$$-$9.4~km/s and at V$_{LSR}$$=$4.4~km/s. The fitted maser positions 
in each velocity channel are listed in Table 3. 
The map was produced from data taken during 
the second observing session.
The maps are in Dec$-$RA offset coordinates,
given relative to the strongest maser feature
at the velocity V$_{LSR}$$=$$-$9.4~km/s $-$ fig. 8.1; on fig. 8.2 positions are given relative to maser feature
at the velocity V$_{LSR}$$=$0.3~km/s (see explanations below).

The maser components given in Table~3 are divided into 4~groups,
according to their velocities.  The 1st and 4th groups are very close
to one another on the sky (see Fig. 8); i.e., {\it dynamically} there
are four groups, but {\it spatially} there are only three groups. 
The central spatial cluster has a spread less than about 20 mas
and includes the 1st and 4th maser groups with velocities in the $-$4.5 to
$+$0.7 km/s range.  Two other spatial clusters $-$ the 2nd and 3rd maser groups $-$
also with spreads each $\sim$20 mas, have velocities in the $-$2.5 to $-$1.2 km/s
range.   Two high-velocity masers are isolated in position and are found
at projected distances of 157 AU and 70 AU, respectively, from the
central spatial cluster.  These three maser clusters fall
in a straight line, about 200 AU in length, as shown in Fig. 8.

In the analysis of \cite{patel} the reference feature for all maser
spots was the high-velocity blue component at V$_{LSR}$$=$$-$14.6~km/s.
This component was detected by \cite{slysh} (albeit with a velocity difference of $\sim$0.5~km/s),
but was not used as their reference maser.  Rather, their reference component was at
V$_{LSR}$$=$$+$1.32~km/s in the central range of the spectrum.  \cite{slysh}
report the relative coordinates of the V$_{LSR}$$=$$-$14.1~km/s component,
which permits us to recalculate the coordinates of
this feature in the reference frame of \cite{patel} and hence to compare
the two observations.
This recalculation shows that
  the differences in the velocities and in the absolute coordinates of
  this component between the two surveys does not exceed the uncertainties
  from both observations: the features observed in \cite{slysh}
  coincide with the group ``B''\ in \cite{patel} (see Fig. 9). According to the
conclusions of \cite{slysh}, 8~maser spots of this group form
 a disk of size 15~AU, while according to \cite{patel} the masers in
the central part of this group form a loop with a
radius of $\leq$1~AU; they suggest that the loop is associated with
a circumstellar shell of outflowing stellar-wind material.

\begin{figure}[H]
\centering
\includegraphics[width=0.76\linewidth]{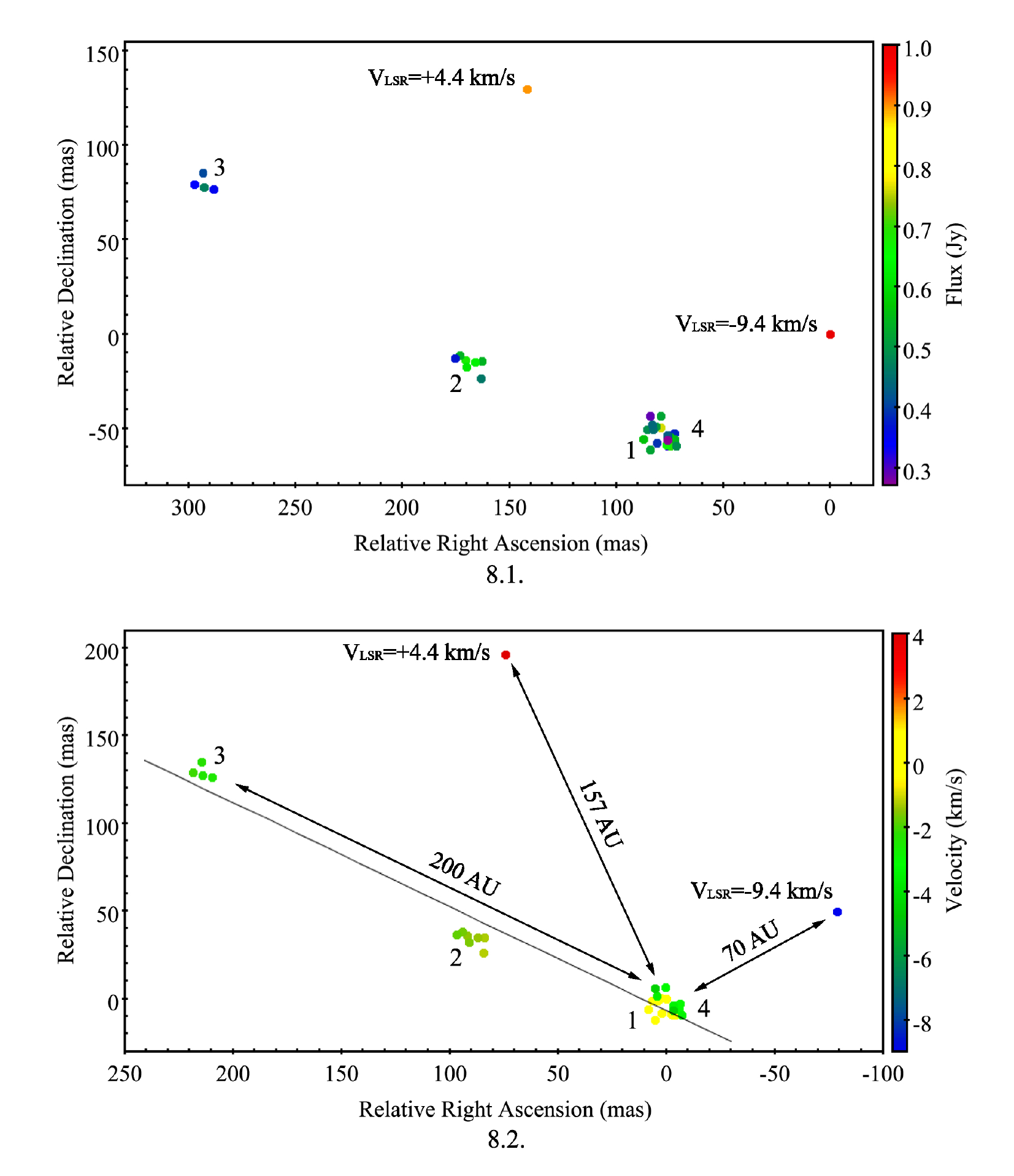} 
\caption{\small{Maps of the H$_2$O maser components in IC~1396~N in relative
Dec$-$RA coordinates for the October session of 2014. 
The central range of the spectrum and two high-velocity components are
presented with the reference at V$_{LSR}$$=$$-$9.4~km/s $-$ Figure 8.1 (a color scale of the 
flux density); and at V$_{LSR}$$=$$+$0.3~km/s $-$ Figure 8.2 (a color scale of the velocity).}}
\end{figure}

The same conversion can be made for the coordinates of the
spectral feature from the reference maser at V$_{LSR}$$=$$-$9.4~km/s, re-scaling
to the frame of the feature at V$_{LSR}$$=$$+$0.3~km/s in our observations (see
Fig.~8). This velocity feature is the only one common
to all three spectra: ours, and those of \cite{patel}
and \cite{slysh}.  By this coordinate conversion we can compare all three data
sets and hence arrive at more robust conclusions.
Although the other features detected in~1996 by \cite{patel, slysh}
were not present at the 2014 epoch, we can determine their positions
with respect to the new velocity components, and thus include these
(past) masers in our analysis.

\begin{figure}[H]
\centering
\includegraphics[width=0.76\linewidth]{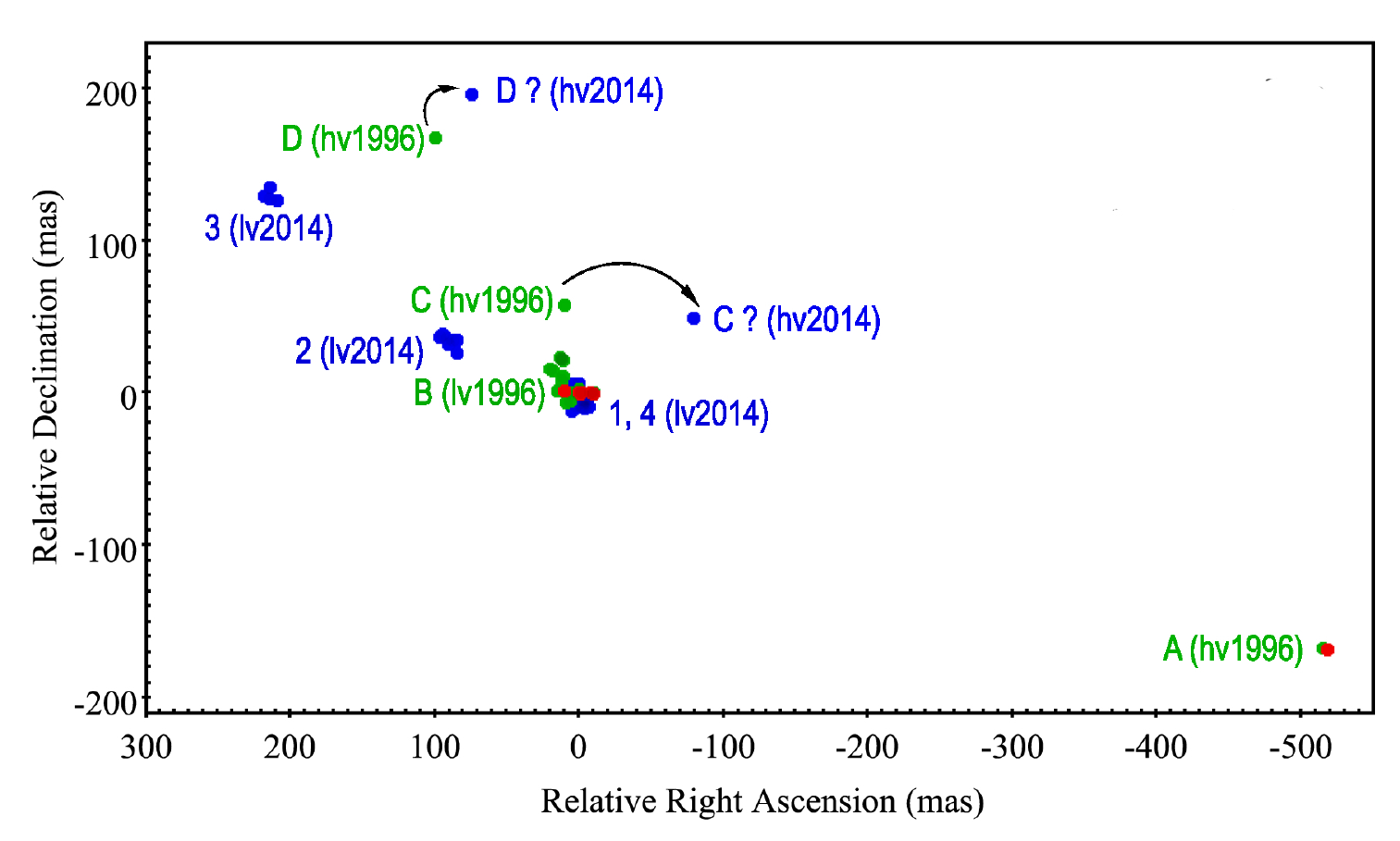}
\caption{\small{Combined map of the H$_2$O maser spots in IC~1396~N
obtained in 1996 and 2014.  The blue circles and numbers on the map correspond to maser groups from Table 3 (2014).  The green circles and letters correspond to maser groups from \cite{patel} (1996).  Also from 1996 are the  red circles from \cite{slysh} (without the distant red feature at velocity
V$_{LSR}$$=$$+$9.3~km/s, which is incompatible with the scale of the map). High-velocity components from the observations of 2014 are indicated with ``hv2014''; high-velocity components from \cite{patel} are indicated by ``hv1996''. Similarly, we denote the low-velocity components as ``lv2014'' (from 2014), and ``lv1996'' (from \cite{patel}). Coordinates were recalculated with the $+$0.3~km/s
feature as the reference position. See description in the text.}}
\end{figure}

In Fig.~9 we show the maser positions of
all three epochs (this paper, \cite{slysh}, \cite{patel}),
calculated relative to the velocity feature at $+$0.3~km/s.
Included in Fig.~9 are the masers from the central part of the
spectrum and also the high-velocity components. The positions from
our observations correspond to the second session of 2014
(including the blue and red features) and are shown as blue circles (numbers on the map correspond to maser groups from Table 3).
The features from \cite{patel}, are
from the central and the more distant ``A position'' (letters in Fig. 9 correspond to maser groups from \cite{patel}) at the
velocity V$_{LSR}$$=$$-$14.6~km/s, and are shown as green circles.
Finally, the masers from \cite{slysh},  including the
high-velocity feature at V$_{LSR}$$=$$-$14.1~km/s but not the most
distant red feature at V$_{LSR}$$=$$+$9.3~km/s, are shown as red circles.

\begin{figure}[H]
\centering
\includegraphics[width=0.76\linewidth]{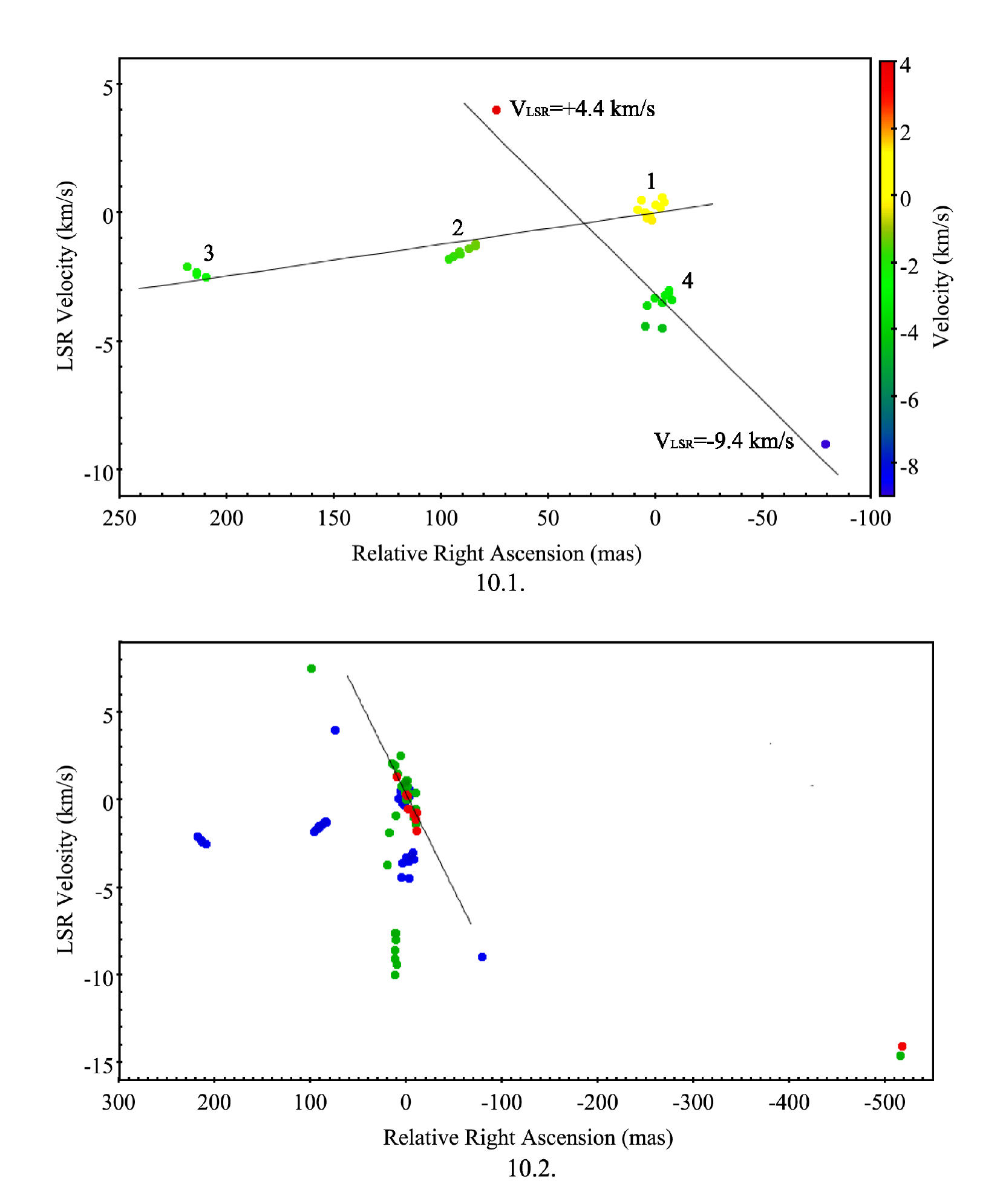}
\caption{\small{Right ascension $-$ velocity diagram for the maser
components of Fig.~8 $-$ corresponds to 10.1~and Fig.~9 $-$ corresponds to 10.2. See description in the text,
Discussion section.}}
\end{figure}

In Fig.~10.1 relative to the velocity
$+$0.3~km/s the Velocity-Right Ascension diagram (V$_{LSR}$-RA) is
shown  for the same session for the central part and two distant
spots V$ _{LSR}$$=$$-$9.4~km/s and V$_{LSR}$$=$4.4~km/s. Two linear
structures are evident:  one consisting of the features with velocities
more positive than V$_{LSR}$$=$$-$3~km/s (the first, second, and third groups;
see Table~3), and a second linear structure consisting of the two
high-velocity components and the fourth maser group centered at velocity
V$_{LSR}$$=$$-$3.4~km/s.

In the combined distribution in  Fig.~9~and~10.2  the same linear
structures are seen.  Also evident is the loop structure observed in 1996
\cite{patel} and the putative Keplerian disk, also observed in 1996
\cite{slysh}, which
corresponds to group one, as observed in 2014 (see Table~3). In
other words, over an 18-year period, the spatial structure
of the maser spots in the central part of the IC~1396~N globule are
confined to the velocity interval $\pm$1.5~km/s around
V$_{LSR}$$=$$+$0.3~km/s, on a scale less than 20~mas or
15~AU.

\subsection{Possible models of the source}

Slysh et al. \cite{slysh} considered a combined model of a Keplerian disk 
surrounding a YSO along with a stellar wind from the central object;
indeed, it is possible that different mechanisms may be
simultaneously involved in the formation of the observed pattern.

Groups 2 and 3, are spatially separated from the central (1,4) cluster and were
first detected only in 2014.  They might be caused by a 
high-velocity jet from the central object/cluster. As seen from
the map (Fig. 8.2), groups 2 and 3 are located almost along a
straight line with respect to the central cluster 
at a distance of 200 AU. The putative jet is collimated to a
narrow opening angle not exceeding 5$^\circ$. These distant
features were not detected in \cite{patel}; see the combined map in Fig. 9.

In \cite{patel} there are two high-velocity features that we did not detect in 2014.
These are shown as green circles in Fig.~9 $-$ one labeled as C (hv1996) and the 
other as D (hv1996).  These two masers may correspond to the two high-velocity
compenents that we mention in the footnote of Table 3.
The presence of high-velocity masers suggests the
action of a collimated outflow from the vicinity of the central
object. The increase of velocity along the line connecting these details could arise from 
the progressive acceleration of the flow by, say, a stellar wind from the central object.
It can be assumed that in the 18 years since the 1996 observations of
\cite{patel} (green circles C(hv1996) and D(hv1996) in Fig.~9) and our observations (blue circles 
C?(hv2014) and D?(hv2014) in Fig.~9) the direction of the outflow has changed by about
$30^\circ$ (see calculations below); this is illustrated by the curved arrows in Fig.~9.

The outflow velocity carrying maser features to a distance
of nearly 200~AU in 18~years must be fairly high $-$ about 56~km/s. This velocity
is sufficient to ionize ambient hydrogen gas and to produce
free-free radio continuum.  There are some centimeter and millimeter continuum studies
of IRAS~21931$+$5802, e.g., at BIMA by \cite{beltran1}.
However, their angular resolution (a few arcseconds) is not high enough to
definitely associate the observed radio structures with the
above-mentioned putative jet exciting the maser features. A higher
resolution radio continuum study of the jet region would be valuable.

A variety of mechanisms have been proposed to explain the
collimation of outflows from  young stellar objects (e.g.,
\cite{rodriguez86, rodriguez99, bachiller, price}): (1)
collimation near the stellar surface related to infalling
circumstellar material; (2) collimation by a circumstellar
structure (a disk or torus); (3) self-focusing of the stellar wind
in the surrounding interstellar medium; and (4)
magnetohydrodynamic mechanisms in the vicinity of the star. In the
case of IC~1396~N, no data on the stellar magnetic field are
available, so nothing can be said about the effect of 
mechanism (4). In particular, no emission from the
magneto-sensitive hydroxyl molecule has been detected.

In view of the evidence for a circumstellar disk in
IC~1396~N, we focus here on mechanism (2). An interesting
phenomenon is the change in direction of the putative outflow with
time.

A possibile cause is the precession of a protoplanetary disk
by the influence of the secondary component of a binary system.
Such a case is discussed in detail, for example, in \cite{lai},
who report an expression for the angular velocity of the disk
precession. Applying their expression, however, yields precession
periods of several thousand years for stellar masses of a few
solar masses and a semi-major orbital axis of a few astronomical
units.

It seems more likely that the collimation and the
precession of the outflow are driven by a larger circumstellar
structure $-$ a torus inside the globule \cite{rudnitskij1, rudnitskij2, berulis}, see Fig.~11. 
This torus is clearly
visible in molecular emission (e.g., N$_2$H$^+$ and CH$_3$CN
\cite{fuente}) as a structure perpendicular to the direction of
the bipolar outflow. Molecular line
data for the circumstellar structure around IRAS~21391$+$5802
\cite{wu} yield a virial mass in the range
of 83$-$186\,M$_{\odot}$. Molecular data indicate that the density
in clumps is decreasing outward as $\rho\propto r^{-\alpha}$, where
$\alpha\sim 1$, see \cite{berulis} and references therein. Adopting
for the circumstellar torus a mass of
134\,M$_{\odot}$ and sizes of 0.06$-$0.14~pc ($\sim 2-4\times
10^{17}$~cm) \cite{wu}, imply that the inner cavity radius 
(see Fig.~11) is
$R_1=H=3\times 10^{15}$~cm, while the ``effective'' outer radius (with
$\alpha =1$) is $R_2 =3\times 10^{16}$~cm.  We then find from the formula
in \cite{rudnitskij1} a disk precession period $T_{\textrm{pr}}\sim
160$~years. Thus, in the 18~years between Patel et al.'s and
our observations the disk could have precessed by about one tenth
of 360$^\circ$, in fair agreement with the $\sim 30^\circ$ we see in Fig.~9.

\begin{figure}[H]
\center{\includegraphics[width=0.4\linewidth]{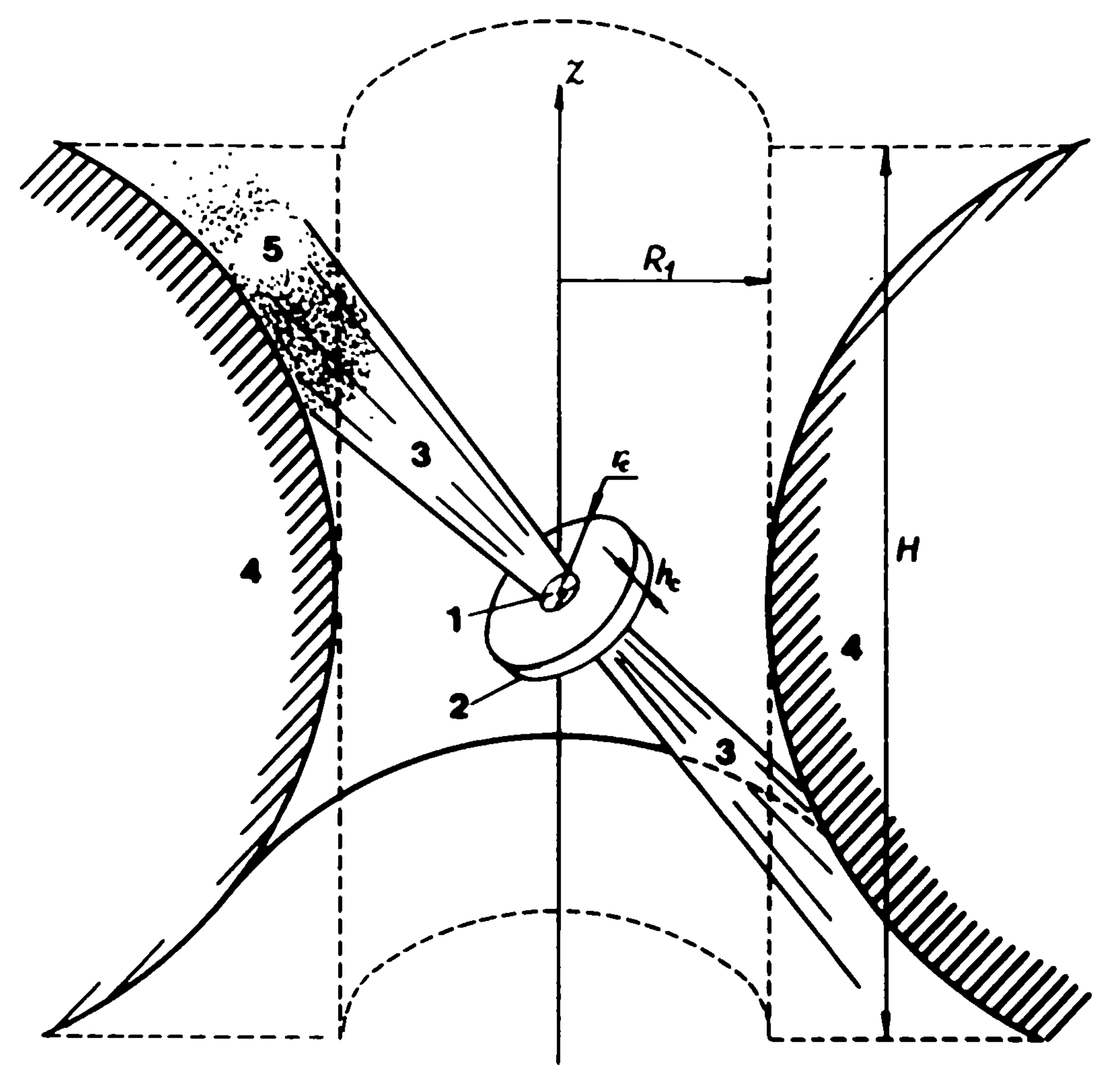}}
\caption{\small{Sketch of the model of the source with a precessing
disk \cite{rudnitskij1}; 1 $-$ star, 2 $-$ protoplanetary disk, 3
$-$ jets, 4 $-$ circumstellar torus, 5 $-$ bright maser spot.}}
\end{figure}

\subsection{Is IC~1396~N losing its protoplanetary disk?}

Another peculiarity of IC~1396~N is a
continuous decline of its H$_2$O maser emission seen in our
Pushchino monitoring during the last 10~years \cite{ash}. Throughout the
first half of the 2000$-$2010 period the maximum H$_2$O line flux density
$F_{\textrm{max}}$ in IC~1396~N was maintained at a level of
hundreds of janskys. On December 19, 2014, the closest date to 
epoch No.~5 of our RadioAstron observations, it was about 50~Jy;
by June 2016 it fell to 20~Jy, and it is continuing to decrease at
the time of submission of this paper. If the observed trend
persists, this may imply the disappearance of the protoplanetary disk
in IC 1396N, which hosted the main features of H$_2$O maser
emission. Such a rapid evolution of the circumstellar structure in
IC~1396~N seems plausible. Rapid changes on a timescale
of a few years were observed in the infrared spectrum of the young
solar-like star TYC~8241~2652~1. From 2008 to 2012 the flux of its
emission between 3 and 300~$\mu$m decreased by a factor of about
30 \cite{melis}. This decrease is interpreted by the loss of the
dust component of the circumstellar protoplanetary disk because of
its concentration into larger bodies (planetesimals). Thus, the
formation of planets in circumstellar disks may be much more rapid
than previously thought.  Perhaps we have managed to catch 
IC~1396~N just at the stage when a drastic restructuring of
the protoplanetary disk results in the decay of H$_2$O maser
emission. A follow-up monitoring program of the H$_2$O maser in IC~1396~N
would confirm or reject this hypothesis.

\section{CONCLUSIONS}

1. During observations of H$_2$O masers in the IC~1396~N globule using
the space-ground interferometer with the 10-m space radio telescope
RadioAstron, no response from the source was detected on
projected baselines $>$2.3~ED. A lower limit on the size of the
masers is L$>$0.03~AU; an upper limit on the brightness
temperature is $T_B=6.25\times10^{12}$~K.

2.  An analysis of the maps obtained by fringe rate mapping for
the session with the longest observing time (4~hours) on
the ground baselines was performed. Parameters of maser spots $-$
positions, velocities, and flux densities $-$ were determined for those
maser features with SNR$>$4.

3. Maser groups at low velocities (in the interval from $-$4.4~km/s to
$+$0.6~km/s) were detected, as were two high-velocity
features at V$_{LSR}$$=$$-$9.4~km/s and V$_{LSR}$$=$$+$4.4~km/s.  The latter two
are spatially offset from the strongest feature  at  $\sim$0.3~km/s
at distances of 157~AU  and  70~AU, respectively.

4. Maser components of the central part of the spectrum can be
divided into four groups based on their velocities. However, two groups (the
first, with a V$_{LSR}$  of  about  $+$0.3 km/s,  and the fourth,
with a V$_{LSR}$ of about $-$3.4~km/s) spatially coincide.

5. On the (RA$-$Dec) maps, four groups of central components
form a chain of maser spots, which is nearly a straight line of
about $\sim$200~AU in extent.  Two groups $-$ the second
($\sim$$-$1.5~km/s) and the third ($\sim$$-$2.4~km/s) $-$ were not
detected in 1996 and perhaps trace a jet that formed during the 18~year
period between 1996 and 2014.

6. The appearance of the high-velocity features ``C'' and ``D'' in 2014 at the different from 1996 \cite{patel} spatial coordinates might be explained by the precession of the collimated outflow under the influence of
gravity from the YSO. 

7. There is a straight line orientation (V$_{LSR}$-distance) seen
in the (V$_{LSR}$-Right Ascension) diagram between the jet and
the first low-velocity group at V$_{LSR}$$\sim$$+$0.3~km/s,
whereas the high-velocity components are aligned with the
4th group at V$_{LSR}$$\sim$$-$3.4.

8. The (V$_{LSR}$-Right Ascension) diagram shows that the
low-velocity maser spots of 1996 and 2014 maintain their positions
near the central velocity V$_{LSR}$$\sim$0.3~km/s during the 18~year
period.

9. The observed continuous decline of the H$_2$O maser emission
in IC~1396~N, especially conspicuous from 2014 to 2016, may indicate
a drastic restructuring of the circumstellar protoplanetary
disk due to the rapid formation of planets within it.

\section{ACKNOWLEDGMENTS}

The RadioAstron project is led by the Astro Space Center of the
Lebedev Physical Institute of the Russian Academy of Sciences and
the Lavochkin Scientific and Production Association under a contract
with the Russian Federal Space Agency, in collaboration with partner
organizations in Russia and other countries.

Partly based on observations with: the 100-m telescope of the MPIfR  (Max-Planck-Institute for Radio Astronomy) at Effelsberg; the Sardinia and Noto Radio Telescopes operated by INAF $-$ Istituto di Radioastronomia; RT-40 telescope in Yebes Observatory operated by IGN $-$ The National Geographic Institute of Spain; the Toru\'n 32-m telescope of the Centre for Astronomy Nicolaus Copernicus University and radio telescopes of IAA RAS (Federal State Budget Scientific Organization Institute of Applied Astronomy of Russian Academy of Sciences).
We are grateful to the
astronomy observatories for the opportunity to observe with the
radio telescopes and to the staff of the observatories for their
assistance in carrying out the observations. 

This work was partially supported by the Russian Foundation for
Basic Research $-$ project 15-02-07676-a (GMR), project 16-32-00877-mol-a (OSB)
and UNAM/DGAPA (project IN114514) (SEK). We thank Dr. S.~V.
Kalenskii for assistance in drafting the observing proposal.


\newpage
\begin{thebibliography}{References}

\bibitem{matthews}
H. E. Matthews,   Astron.  and  Astrophys. \textbf{75},  345 (1979).

\bibitem{fuente}
A. Fuente,   T.  Castro-Carrizo, T. Alonso-Albi,  M. T. Beltr\'{a}n,
C.  Ceccarelli,  B.  Lefloch, C.  Codella,  and P. Caselli, Astron.
and  Astrophys. 507,  1475  (2009).

\bibitem{beltran1}
M. T. Beltr\'{a}n,  J. M.  Girart,  R. Estalella,  P. T. P. Ho, and
A. Palau,  Astrophys.  J.  573, 246 (2002).

\bibitem{codella}
C. Codella, R. Bachiller, B. Nisini, P. Saraceno, and L. Testi,
Astron. and Astrophys. 376, 271, (2001).

\bibitem{sugitani}
K. Sugitani,  Y. Fukui,  A. Mizuni,  and  N. Ohashi,  Astrophys.  J.
\textbf{342},  L87  (1989).

\bibitem{wilking} B. A. Wilking, J. H. Blackwell, and L. G. Mundy,
    Astron.  J.  \textbf{100}, 758 (1990).

\bibitem{gyulbudaghian}
A. L. Gyulbudaghian,  L. F. Rodr\'{\i}guez,  and  S.  Curiel,
Rev. Mex. Astron. y Astrof{\'\i}s.   \textbf{20},  51
(1990).

\bibitem{edris} K.~A.~Edris, G.~A.~Fuller, and R.~J.~Cohen,
    Astron. and Astrophys. 465, 865 (2007).

\bibitem{slysh67} V. I. Slysh, I. E. Val'tts, S. V. Kalenskii, M. A. Voronkov, F. Palagi, G. Tofani, M. Catarzi, Astron. and Astrophys. Suppl. Ser. \textbf{134}, 115 (1999).
		
\bibitem{ic44ghz} S. V. Kalenskii, R. Bachiller, I. I. Berulis, I. E. Valtts, J. Gomez-Gonzalez, J. Martin-Pintado, A. Rodriguez-Franco, V. I. Slysh, Soviet Astronomy. \textbf{36}, 517 (1992).

\bibitem{irdc} R. Simon, J. M. Jackson, J. M. Rathborne, and
E. T. Chambers, Astrophys. J. \textbf{639}, 227 (2006).

\bibitem{sdc} N. Peretto and G. A. Fuller, Astron. and Astrophys.
\textbf{505}, 405 (2009).

\bibitem{ego} C. J. Cyganowski, B. A. Whitney, E. Holden,
E. Braden, C. L. Brogan, E. Churchwell, R. Indebetouw,
D. F. Watson, B. L. Babler, R. Benjamin,
M. Gomez, M. R. Meade, M. S. Povich, T. P. Robitaille,
and C.Watson, Astron. J. \textbf{136}, 2391 (2008).

\bibitem{tofani} G. Tofani,  M. Felli,  G. B. Taylor,  and  T. R.
    Hunter,  Astron. and Astrophys. Suppl.  Ser. 112,  299 (1995).

\bibitem{migenes}
V. Migenes, S. Horiuchi,  V. I. Slysh,  I. E. Val'tts, V.  Golubev,
P. G. Edwards, E. B.  Fomalont,  R. Okayasu, P. J.  Diamond, T.
Umemoto,  and M.  Inoue, Astrophys.  J.  Suppl.  Ser. 123,
487  (1999).

\bibitem{slysh}
V. Slysh,  I. Val'tts,  V. Migenes,  E. Fomalont,  H. Hirabayashi,
M.  Inoue, and T. Umemoto, Astrophys.  J.  526, 236 (1999).

\bibitem{patel} N. A. Patel,  L. J. Greenhill, J. Herrnstein,
    Q.~Zhang, J.~M.~Moran, P.~T.~P.~Ho, and
    P.~F.~Goldsmith,  Astrophys. J. 538,  268  (2000).

\bibitem{rudnitskij1}
G. M. Rudnitskij,  1987a, Proceedings of IAU Symp. No. 115, held
11-15 November 1985 in Tokyo, Japan.  M. Peimbert and J. Jugaku,
eds. Dordrecht: D. Reidel Publishing Co., p.  398 (1987a).

\bibitem{rudnitskij2}
G. M. Rudnitskij,  Bull. Astron. Inst. Czechoslovakia  \textbf{69}, 51 (1987b).

\bibitem{berulis}
I. I. Berulis,  E. E.  Lekht,  and  G. M. Rudnitskij,  Astron. Rep. \textbf{40}, 36 (1996).

\bibitem{lightfoot}
J. F. Lightfoot,  Monthly Notices Roy. Astron. Soc.  \textbf{239}, 665  (1989).

\bibitem{kardashev}
N.~S.   Kardashev,  V.~V.   Khartov,    V.~V.   Abramov,   V.~Yu.
Avdeev,        A.~V.   Alakoz,    Yu.~A.   Aleksandrov,     S.
Ananthakrishnan,     V.~V.   Andreyanov,      A.~S. Andrianov, N.~M.
Antonov,     M.~I.   Artyukhov,    W.   Baan,    N.~G. Babakin,
V.~E.   Babyshkin,     K.~G.   Belousov,     A.~A. Belyaev, J.~J.
Berulis,      B.~F.   Burke,      A.~V. Biryukov,      A.~E. Bubnov,
M.~S.   Burgin,    G.   Busca, A. A.   Bykadorov, V.~S.   Bychkova,
V.~I.   Vasil'kov, K.~J.   Wellington, I.~S.   Vinogradov,     R.
Wietfeldt,   P. A.   Voitsik,    A.~S. Gvamichava,    I.~A.   Girin,
L.~I. Gurvits,      R.~D. Dagkesamanskii,      L.   D'Addario, G.
Giovannini,       D.~L. Jauncey,      P.~E.   Dewdney,     A.~A.
D'yakov,      V.~E. Zharov,    V.~I.   Zhuravlev,     G.~S.
Zaslavskii,     M.~V. Zakhvatkin,      A.~N.   Zinov'ev, A.~V.
Ipatov,      B.~Z. Kanevskii,      I.~A.   Knorin, J.~L.   Casse,
K.~I. Kellermann,      Yu.~A.   Kovalev, Y.~Y.   Kovalev,      A.~V.
Kovalenko,     B.~L.   Kogan, R.~V.   Komaev,      A.~A.
Konovalenko,      G.~D.   Kopelyanskii,    Yu.~A.   Korneev, V.~I.
Kostenko,     B.~B.   Kreisman, A.~Yu.   Kukushkin, V.~F.
Kulishenko,   D.~N.   Cooper, A.~M.   Kut'kin,    W.~H. Cannon,
M.~G.   Larionov,     M.~M. Lisakov,      L.~N. Litvinenko, S.~F.,
Likhachev, L.~N.   Likhacheva,    A.~P. Lobanov,     S.~V.
Logvinenko, G.   Langston,    S.~Yu. Medvedev,    M.~V.   Melekhin,
A.~V. Menderov,      D.~W. Murphy,     T.~A.   Mizyakina,     Yu.~V.
Mozgovoi,      N.~Ya. Nikolaev,      B.~S.   Novikov,     I.~D.
Novikov,     V.~V. Oreshko,    Yu.~K.   Pavlenko,     I.~N.
Pashchenko,     Yu.~N. Pomomarev,   M.~V.   Popov,    A.
Pravin-Kuma,     R.~A. Preston,     V.~N.   Pyshnov,     I.~A.
Rakhimov,     V.~M. Rozhkov,   J.~D.   Romney,      P.   Rocha,
V.~A.   Rudakov,     A. R\"ais\"anen,    S.~V.   Sazankov, B.~A.
Sakharov,     S.~K. Semenov,     V.~A.   Serebrennikov, R.~T.
Schilizzi,     D.~P. Skulachev,      V.~I.   Slysh, A.~I.   Smirnov,
J.~G.   Smith, V.~A.   Soglasnov, K.~V.   Sokolovskii,      L.~H.
Sondaar, V.~A.   Stepan'yants, M.~S.   Turygin,     S.~Yu. Turygin,
A.~G.   Tuchin,    S. Urpo,    S.~D.   Fedorchuk,     E.~B.
Fomalont,     I.~ Fejes, A.~N.   Fomina,     Yu.~B.   Khapin, G.~S.
Tsarevskii, J.~A.   Zensus,    A.~A.   Chuprikov,     M.~V.
Shatskaya, N.~Ya.   Shapirovskaya,    A.~I.   Sheikhet,     A.~E.
Shirshakov, A.   Schmidt,     L.~A.   Shnyreva,    V.~V.
Shpilevskii, R.~D.   Ekers,    V.~E.   Yakimov, Astron. Rep.
\textbf{57}, 153 (2013).

\bibitem{kov14}
Yu. A. Kovalev, V. I. Vasil'kov, M. V. Popov, V. A. Soglasnov, P. A.
Voitsik, M. M. Lisakov, A. M. Kut'kin, N. Ya., N. A. Nizhel'skii, G.
V. Zhekanis, P. G. Tsybulev,  Cosmic Research \textbf{52}, 393
(2014).

\bibitem{andr14}
A. S. Andrianov, I. A. Girin, V. E. Jharov, V. I. Kostenko, S. F.
Likhachev, and M. V. Shatskaya,  Vestnik NPO im. S. A. Lavochkina
No. 3, 55 (2014).

\bibitem{bachiller} R.~Bachiller, Ann. Rev. Astron. and
    Astrophys. 34, 111 (1996).

\bibitem{rodriguez86} L.~F.~Rodr{\'\i}guez, Publ. Astron. Soc.
    Pacific. 98, 1012 (1986).

\bibitem{rodriguez99} L.~F.~Rodr{\'\i}guez, in \emph{Proceedings
    of Star Formation 1999, held in Nagoya, Japan, June 21$-$25,
    1999}, Ed. T.~Nakamoto, (Nagoya: Nagoya Univ., 1999), p.~257.

\bibitem{price}
D.~J.~Price, T.~S.~Tricco, and M.~R.~Bate,
    Monthly Notices Roy. Astron. Soc. 423, L45 (2012).

\bibitem{lai}
D.~Lai, Monthly Notices Roy. Astron. Soc. 440, 3532 (2014).

\bibitem{wu}
J.~Wu, N.~J.~Evans II, Y.~L.~Shirley, and C.~Knez, Astrophys. J.
    Suppl. Ser. \textbf{188}, 313 (2010).

\bibitem{ash} N.~Ashimbaeva , O.~Bayandina, P.~Colom,
E.~Lekht, M.~Pashchenko, G.~Rudnitskij, A.~Tolmachev, I.~Val'tts,
talk presented at Symposium No.~9, European Week of Astronomy and
Space Science, Athens, Greece, 4--8~July 2016.

\bibitem{melis} C.~Melis, B.~Zuckerman, J.~H.~Rhee,
I.~Song, S.~J.~Murphy, and M.~S.~Bessell, Nature \textbf{487}, 74
(2012).

\end{thebibliography}
\end{document}